# Density Field Reconstruction of an Overexpanded Supersonic Jet using Tomographic Background-Oriented Schlieren

Joachim A. Bron[1]*, Woutijn J. Baars[2], Ferdinand F. J. Schrijer[2]

August 2023

Faculty of Aerospace Engineering
Delft University of Technology

**Abstract**

A Tomographic Background-Oriented Schlieren (TBOS) technique is developed to aid in the visualization of compressible flows. An experimental setup was devised around a sub-scale rocket nozzle, in which four cameras were set up in a circular configuration with 30° angular spacing in azimuth. Measurements were taken of the overexpanded supersonic jet plume at various nozzle pressure ratios (NPR), corresponding to different flow regimes during the start-up and shut-down of rocket nozzles. Measurements were also performed for different camera parameters using different exposure times and f-stops in order to study the effect of measurement accuracy. Density gradients and subsequently two-dimensional line-of-sight integrated density fields for each of the camera projections are recovered from the index of refraction field by solving a Poisson equation. The results of this stage are then used to reconstruct two-dimensional slices of the (time-averaged) density field using a tomographic reconstruction algorithm employing the filtered back-projection and the simultaneous algebraic reconstruction technique. By stacking these two-dimensional slices, the (quasi-) three-dimensional density field is obtained. The accuracy of the implemented method with a relatively low number of sparse cameras is briefly assessed and basic flow features are extracted such as the shock spacing in the overexpanded jet plume.

# Graphical abstract

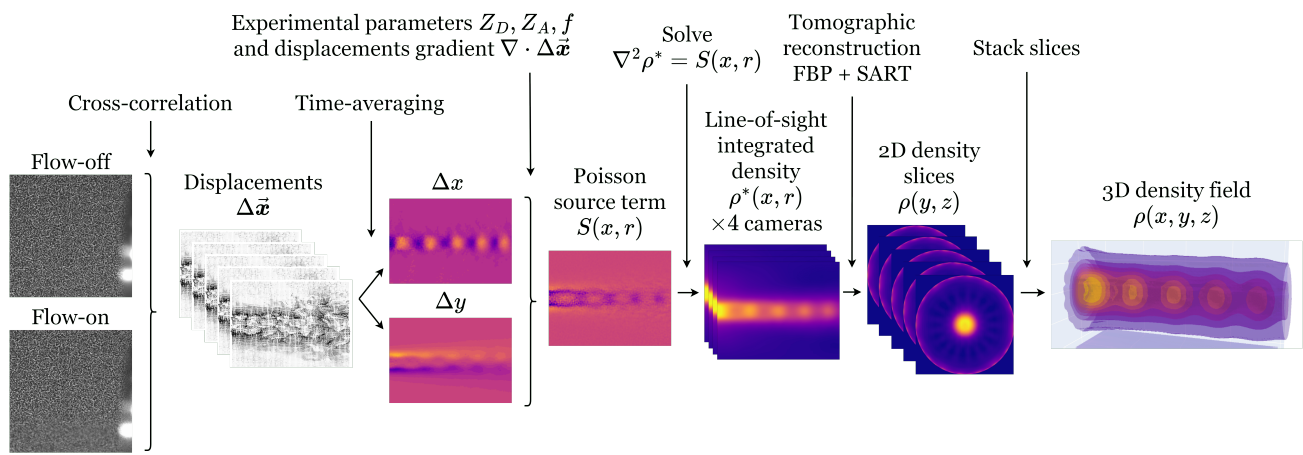

*Corresponding author: ✉ joabron@mit.edu
[1] B.Sc. student, Faculty of Aerospace Engineering, Delft University of Technology, The Netherlands
[2] Department of Flow Physics and Technology, Faculty of Aerospace Engineering, Delft University of Technology, The Netherlands

# 1 Introduction

THE Background-Oriented Schlieren (BOS) technique is a flow visualization technique used to visualize density gradients in compressible flows. It uses the deflection of light rays due to refractive index variations in a medium, caused by changes in the medium's density. Using digital correlation techniques similar to those used in particle image velocimetry (PIV), the apparent displacements of a background pattern can be computed, which is directly proportional to the density gradients in the flow. It is of the same family as classical Schlieren photography, interferometry and shadowgraphy (see Settles 2001), since these techniques use density gradients of the flow or derivatives hereof.

The method was initially described by Dalziel et al. (2000) under the name of "synthetic Schlieren", and by Raffel et al. (2000a) as a qualitative tool for visualizing helicopter blade tip vortices. Early publications by Meier (2002) and by Richard et al. (2002) showed the potential of the BOS technique as applied to a variety of different applications. These ranged from the density field of supersonic jets to determining concentrations of gas mixtures. These publications also mentioned the potential for a combination of BOS with tomographic techniques to retrieve the temperature, pressure and density fields of flows. Some of the practical aspects of performing BOS measurements were laid out by Richard and Raffel (2001).

Although early publications focused on the practical aspects of the method and its potential, the results were mostly qualitative. Early literature often mentioned the possibility of combining BOS with tomography, but it was not until later efforts that quantitative field reconstructions were demonstrated. Venkatakrishnan and Meier (2004) showcased the first conjunction of BOS with tomography, also called Tomographic Background-Oriented Schlieren (TBOS), applied to an experimental Mach 2 flow around a cone and were able to accurately reconstruct the three-dimensional density field. However, use was made of an axis-symmetric assumption, meaning the TBOS used is not the most general and can not be used for non-axis-symmetric cases. This assumption was not made by Goldhahn and Seume (2007), who were able to reconstruct the three-dimensional density field of an under-expanded double-orifice jet using the filtered back projection (FBP) algorithm and 36 projections. Their results show that the reconstructed field, where shock diamonds are clearly visible, was obtained with good resolution. As opposed to Venkatakrishnan and Meier (2004), their method does not use the Poisson equation but uses the density gradients directly in the tomographic reconstruction. Additionally, the findings show that the focal length and camera resolution are the factors most affecting the sensitivity and resolution of the reconstruction, as well as the relative distance between object, background, and camera. Surprisingly, they found that the overall length of the setup plays a minor role in the sensitivity.

Grauer et al. (2018) presented the first application of BOS tomography to the reconstruction of a combustion process. A 23 camera setup was used to perform BOS measurements, and tomographically combine these various projections to reconstruct the 3D instantaneous refractive index distribution of an unsteady natural air/gas flame. They use a slightly different approach with respect to the tomography than the two previously mentioned works. Their method is based on a Bayesian framework using total variation (TV) priors, which was shown to be well suited for strong gradients as the authors were able to properly reconstruct the abrupt changes in refractive index. Supersonic jet applications could potentially benefit from this method, since abrupt refractive index changes occur due to the strong density gradients of the shear layers and shock-waves.

Raffel (2015) presents a comprehensive review of recent advances in BOS imaging, its applications and variations of the technique. He shows that the advantages of the BOS technique over other techniques such as classical Schlieren are predominantly its experimental simplicity (BOS requires only a camera and background pattern with enough spatial frequency) and the robustness of the digital correlation algorithms, which are well established as these are widely used in techniques such as PIV (an extensive treatment of PIV is given by Raffel et al. 2007). BOS's main disadvantage is the inherent limited resolution due to the displacement computation being averaged over so called interrogation windows. Furthermore, the camera is typically focused on the background to obtain good contrast, which inherently limits the resolution of the flow being studied which is out of focus. In recent years, the BOS technique has been applied more often (see Raffel 2015) due to the advent of digital correlation techniques and better resolution cameras.

The simplicity of the experimental setup leads to another advantage of the BOS technique over other flow visualization techniques, which is its capability of measuring large flow fields. Doing this with techniques such as classical Schlieren or interferometry would be difficult, as these require more optical hardware components such as large mirrors. An early demonstration of the capability of applying BOS to large scale flows was done by Raffel et al. (2000b), where the technique was applied to visualize helicopter blade tip vortices using the natural environment as a background. More recently, Heineck et al. (2021) used BOS imaging on a full-scale supersonic aircraft in flight, using the desert fauna as a background pattern, and were able to produce the most accurate density gradient images of aircraft in flight to date (see fig. 1).

Although there has been much development in the field, there have not been many studies performed that use few projections in the TBOS reconstruction. To the best of the author's knowledge, no study has been performed that implemented TBOS using less than 5 projections. Implementing TBOS using few-projections is challenging as the tomography, which is an inverse problem, needs as many projections as possible for proper flow field reconstruction.

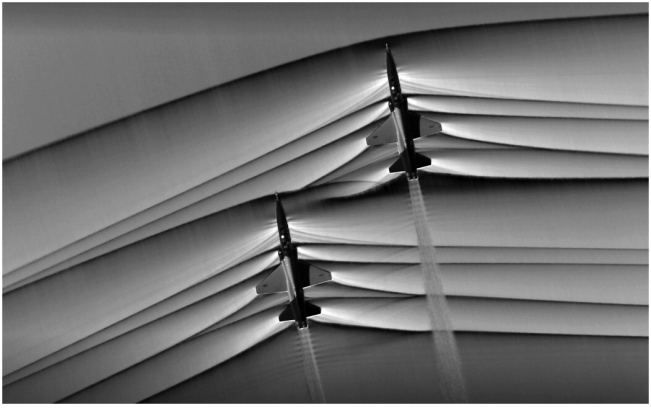

Figure 1: Airborne BOS imaging of a formation of two T-38 aircraft at $M = 1.02$ reveals shockwaves (from Heineck et al. 2021)

The development of a robust and accurate few-projections TBOS technique presents a real opportunity, especially in situations where only a limited number of viewing angles are present and/or possible. This is the case, for example, in supersonic wind tunnels where not many viewing angles are possible. In the present study, TBOS will be applied to the over-expanded supersonic jet of a sub-scale rocket nozzle, with the goal of reconstructing the flow's three-dimensional density field. This application was chosen as it serves a good test case: it is an external flow, meaning there are no problems placing the cameras, it has stream-wise development (shocks), and it is of practical relevance. Furthermore, since the transient flow characteristics occurring during the start-up and shut-down phases are difficult to predict, current nozzles are over-designed to withstand the critical lateral forces that occur during these phases (Frey and Hagemann 1999; Baars et al. 2012; Östlund et al. 2004). Applying the TBOS technique to rocket nozzles could, in the future, aid in the understanding of the side loads that occur during these phases, and through this help optimize their design and reduce their mass, directly translating to higher payload capabilities.

In section 2, the objectives of the present study will be laid out. Then, in section 3, the theoretical background of the BOS technique and the principle of tomographic reconstruction will be presented. Section 4 gives a detailed description of the experimental method and data acquisition, followed by the data processing steps given in section 5. Based on these, the results are obtained and analyzed in section 6. Finally, section 7 gives some concluding remarks and future recommendations.

## 2 Objectives and contribution

To reconstruct the jet's density field, the following approach will be taken. BOS will be used to extract the density gradients at various projections orthogonal to the main nozzle axis. Using these projections, a tomographic algorithm will be used to reconstruct individual slices of the flow (similar to Venkatakrishnan and Meier 2004), and by stacking these, the full three-dimensional field will be reconstructed.

The main objective of the present study is to develop an experimental setup, data acquisition configuration, and collection of data post-processing scripts as an initial framework for performing TBOS research on jet plumes at the Delft University of Technology Aerodynamics High-Speed Laboratory (HSL). Additionally, the goal is to explore the feasibility of doing so using a low number of sparsely placed cameras. Hereafter, experimental BOS data will be generated of a supersonic (overexpanded) sub-scale rocket nozzle jet at various nozzle pressure ratio (NPR) and camera parameters, and the developed TBOS technique will be applied to this data to create quantitative, three-dimensional reconstructions of the density field.

After this initial iteration, the aspects of this framework could in the future be extended and improved, and could perhaps even be used as a starting point for other tomographic flow measurement techniques. Finally, this paper serves as an initial reference point for some important TBOS literature (specifically applied to rocket nozzle jets), and recommendations for TBOS experimental design at the HSL.

## 3 Theoretical background

### 3.1 BOS working principle

The following subsection largely follows the review by Raffel (2015). The basis for the BOS method is based on a medium's change in refractive index caused by density gradients (Richard et al. 2002). A medium's refractive index is defined as

$$n = \frac{c_0}{c} \quad (1)$$

where $c_0$ is the speed of light in vaccuum and $c$ is the speed of light in the medium. When light encounters a denser and thus more refractive medium, it travels slower. Thereby, if there is a gradient in density and thus refractive index, the wavefront will tilt and the ray will be refracted. The relation between a gaseous medium's refractive index and its density is given by the Gladstone-Dale equation:

$$n = 1 + G(\lambda)\rho \quad (2)$$

where $n$ is the refractive index of the medium, $\rho$ is the density of the medium, and $G$ is the Gladstone-Dale constant, given by

$$G(\lambda) = 2.2244 \times 10^{-4} \left[1 + \left(\frac{6.7132 \times 10^{-8}}{\lambda}\right)^2\right]. \quad (3)$$

Here, $\lambda$ is the wavelength of light used, and is taken (for white light) as $\lambda \approx 550$ nm, resulting in $G = 2.26 \times 10^{-4} \, \text{m}^3 \, \text{kg}^{-1}$.

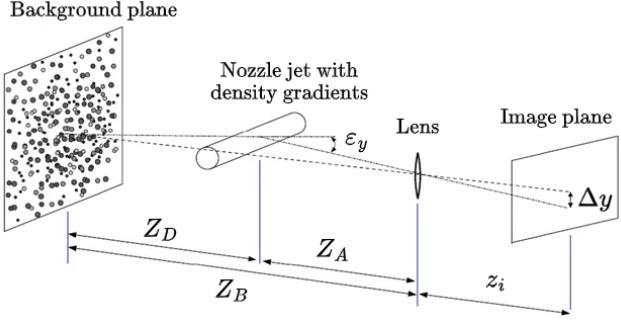

Figure 2: The BOS setup (adapted from Raffel 2015)

An overview of the BOS technique setup is given in fig. 2. The simple setup consists of a background pattern with high spatial frequency (usually random dots or wavelet noise), the medium or flow with density gradients to be studied, and an imaging device, usually a camera. By comparing recordings of the background with flow present ("flow on") and no flow ("flow off"), an apparent shift in the background can be extracted. This apparent shift in the background is due to the light rays being refracted when the flow is present.

The image displacement and light ray deflection at a point in the image plane are defined respectively as

$$\Delta \vec{x} = \begin{bmatrix} \Delta x & \Delta y \end{bmatrix}^T \quad (4)$$

$$\vec{\varepsilon} = \begin{bmatrix} \varepsilon_x & \varepsilon_y \end{bmatrix}^T \quad (5)$$

where $\Delta x$ and $\Delta y$ are the displacements and $\varepsilon_x$ and $\varepsilon_y$ are the deflections, in the $x$ and $y$ coordinates of the image, respectively. Assuming small angles ($\tan \varepsilon \approx \varepsilon$), the image displacement can be written as

$$\Delta \vec{x} = Z_D M \vec{\varepsilon} \quad (6)$$

where $M = z_i/Z_B$ is the magnification factor of the background, $Z_D$ is the distance between the medium and the background pattern, $Z_B$ is the distance between the camera lens and the background pattern, and $z_i$ is the distance between the lens and the image plane (usually taken as the focal length of the lens used). Furthermore, light coming from the background contains information on the refractive index gradients integrated along the line-of-sight, and the deflection of a light ray is given by:

$$\vec{\varepsilon} = \frac{1}{n_0} \int_{z_D - \Delta z_D}^{z_D + \Delta z_D} \nabla n \, dz. \quad (7)$$

where $\nabla n = \begin{bmatrix} \frac{\partial n}{\partial x} & \frac{\partial n}{\partial y} \end{bmatrix}^T$. Based on the (thin) lens equation, it can be shown that $M = f/(Z_B - f)$, and the image displacement can be rewritten as

$$\Delta \vec{x} = f \left( \frac{Z_D}{Z_D + Z_A - f} \right) \vec{\varepsilon} \quad (8)$$

where $Z_A$ is the distance between the camera lens and the medium, and $f$ is the camera's focal length. For best contrast of the background and to ensure a proper extraction of the image displacements using cross-correlation, the camera needs to be focused on the background. With the background in focus, we have the following:

$$\frac{1}{f} = \frac{1}{z_i} + \frac{1}{Z_B}. \quad (9)$$

However, the sharp imaging of the density gradients of the medium would be best if the setup was focused on the flow, leading to

$$\frac{1}{f} = \frac{1}{Z_i'} + \frac{1}{Z_B}. \quad (10)$$

where $Z_i'$ is the distance between the lens and image plane for the flow to be in focus. This focusing problem is an inherent challenge in BOS imaging. The geometric or optical blur $d_i$ of a point at $Z_A$ in the medium is given by:

$$d_i = \frac{1}{f_\#}[f - M'(Z_A - f)] \quad (11)$$

with $M' = Z_i'/Z_A$ the magnification factor of the flow, and $f_\# = f/d_A$ the f-number or f-stop of the camera, and $d_A$ the aperture diameter. According to eq. (8), to achieve a high background displacement a large distance between the flow and background $Z_D$ is needed, and the Schlieren object needs to be placed close to the camera (small $Z_A$). However, since the camera is focused on the background, the closer the Schlieren object is to the camera, the more out of focus the density gradients, and the higher the optical blur (eq. 11). This is an inherent trade-off in BOS imaging. On top of optical blurring, the imaging of small scale structures on the background is also diffraction limited, and the diffraction-limited minimum diameter is given by

$$d_d = 2.44 f_\#(M + 1)\lambda. \quad (12)$$

This equation also gives information on the maximum allowable $f_\#$ to avoid peak locking (Michaelis et al. 2016) based on a certain speckle diameter. To optimize the overall sharpness of the BOS setup, the overall image blur needs to be minimized. The overall image blur is given by the following approximation:

$$d_\Sigma = \sqrt{d_i^2 + d_d^2} \quad (13)$$

Clearly, larger apertures and thus smaller $f_\#$ will increase the geometric blur, but decrease the diffraction limited diameter. However, according to Raffel (2015), geometric blurring usually has a bigger effect on the overall blur, and thus to minimize it large f-numbers and thus small aperture diameters are used, which increases the need for intense background illumination. Furthermore, as long as the overall image blur is much smaller than the window size ($d_\Sigma \ll WS$), there is no significant loss of information since the correlation algorithm averages over the image sub-regions.

According to Venkatakrishnan and Meier (2004), the derivative of the density gradients leads to the Poisson equation:

$$\frac{\partial^2}{\partial x^2}\rho^*(x, y) + \frac{\partial^2}{\partial y^2}\rho^*(x, y) = S(x, y) \quad (14)$$

where $\rho^*(x, y)$ is the line-of-sight integrated density, and where the source term $S(x, y)$ is proportional to the derivative of the displacements obtained through BOS and a scaling factor depending on the optical system. More specifically, $S$ at each point in the image is given as:

$$S = \frac{Z_B}{Z_D f} \frac{n_0}{G} \left( \frac{\partial \Delta x}{\partial x} + \frac{\partial \Delta y}{\partial y} \right) \quad (15)$$

This equation can be solved for the line-of-sight integrated density field, $\rho^*(x, y)$, which can then be used in combination with a tomographic reconstruction to recover the three-dimensional density field. Details on the numerical implementation are delayed until section 5.

## 3.2 Cross-correlation

To extract the pixel displacements of the apparent shift in background by comparing "flow off" and "flow on" images, use is made of the cross-correlation algorithm commonly used in PIV. The algorithm consists of first splitting both flow-off and flow-on images into sub-regions of a certain size called interrogation window (IW). Corresponding IWs of the flow-on and flow-off images are then cross-correlated, leading to a correlation map where the peak corresponds to the most likely average displacement of the background pattern in the IW between the flow-off and flow-on images. This operation is performed for all IWs of the flow-off and flow-on images, leading to a map where at each IW's center location the best guess displacement in $x$ and $y$ is given.

In this study, use is made of the zero-normalized cross correlation as it is unaffected by changes in intensity of the images. This operation results in a correlation coefficient function given by (Raffel et al. 2007)

$$\gamma(\Delta x, \Delta y) = \frac{\sum_{x,y}^{W} \left[ I_a(x,y) - \bar{I}_a \right] \left[ I_b(x + \Delta x, y + \Delta y) - \bar{I}_b \right]}{\sqrt{\sum_{x,y}^{W} \left[ I_a(x,y) - \bar{I}_a \right]^2 \sum_{x,y}^{W} \left[ I_b(x,y) - \bar{I}_b \right]^2}} \quad (16)$$

where $I_a$ and $I_b$ are the pixel intensities, $\bar{I}_a$ and $\bar{I}_b$ are the average pixel intensities over the window, $W$ is the window size, and $\Delta x$ and $\Delta y$ are the window shifts. The most probable displacements in $x$ and $y$ of an IW is given by finding the position of the peak in the correlation map:

$$\Delta \vec{x} = \underset{\Delta x, \Delta y}{\arg \max}\, \gamma(\Delta x, \Delta y) \quad (17)$$

The cross-correlation inherently reduces the resolution since the displacements are averaged for all the pixels in the IW. This can be partially mitigated by using an overlap factor, at the cost of an increase in computation time. For faster computations, use is made of a fast fourier transform (FFT) implementation of the cross-correlation (Lewis 1995). Note that to obtain a high signal-to-noise ratio (SNR) correlation peak and thus a high quality displacement guess, the background pattern must have enough contrast and high spatial frequency. The optimal size of the speckles is between 2-3 pixels (Vinnichenko et al. 2012; Scharnowski and Kähler 2020), but usually only speckles of 3-5 pixels can be attained.

## 3.3 Tomographic reconstruction

The general idea behind tomographic reconstruction is to reconstruct a certain field from its projections which have one less dimension. In this study, the goal is to reconstruct the three-dimensional density field using two-dimensional projections of this field, which are related to the BOS displacements. This is done by so called "back-projecting" the projections to obtain an estimate of the original field. This general idea is shown in fig. 3.

According to Feng et al. (2002), a projection $P_\theta(t)$ at an angle $\theta$ is given by:

$$\begin{aligned} P_\theta(t) &= \int_{\text{ray}} \rho(x, y) \mathrm{d}l \\ &= \iint_{\mathbb{R}^2} \rho(x, y) \delta(x \cos \theta + y \sin \theta - t) \mathrm{d}x\, \mathrm{d}y \end{aligned} \quad (18)$$

Based on the earlier introduced notation, we have that $\rho^*(x, y)$ (at a certain projection angle) corresponds to the projection $P_\theta(t)$. Furthermore, using the Fourier transform and Fourier slice theorem, it can be shown that (see also Venkatakrishnan and Meier 2004)

$$\rho(x, y) = \int_0^\pi \int_{-\infty}^\infty P_\theta(\zeta) h(t - \zeta) \mathrm{d}\zeta \mathrm{d}\theta \quad (19)$$

where

$$h(t) = \int_{-\infty}^\infty H(\omega) e^{j\omega t}\, \mathrm{d}\omega \quad (20)$$

$$H(\omega) = \begin{cases} \pi |\omega| \operatorname{sinc}(\omega/\omega_{\max}), & |\omega| \leqslant \omega_{\max} \\ 0, & |\omega| > \omega_{\max} \end{cases} \quad (21)$$

is the Shepp-Logan filter, which is better at dealing with high frequency noise than the ramp filter. In this study, individual slices orthogonal to the nozzle axis are reconstructed, and these two-dimensional density slices are stacked together to reconstruct the (quasi-) three-dimensional density field. It should noted that although here the integrated line-of-sight density is back projected, it is also possible to back-project the density gradients and only then solve the Poisson equation, as done by Ichihara et al. (2022) and Kirby et al. (2017).

It is important to note that here use is made of the parallel beams assumption, which provides a good approximation for the reconstruction. A more accurate direct fully three-dimensional reconstruction method using Bayesian statistics can be implemented which does not make use of this assumption, but is more computationally demanding. The reader is referred to Grauer et al. (2018), Nicolas et al. (2017) and Amjad et al. (2020).

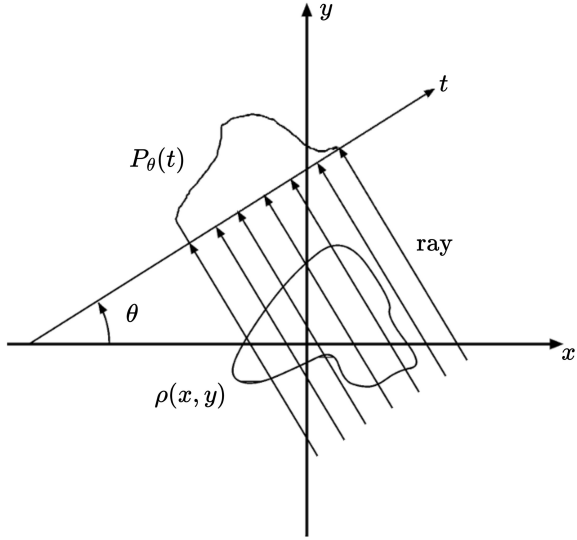

Figure 3: Relation between projections and main field in computed tomography (adapted from Feng et al. 2002)

Table 1: Optical parameters and equipment used

| Parameter | Description |
|---|---|
| Camera model | 4× Imperx Bobcat IGV-B1610 |
| Image resolution | 1624 × 1236 pix |
| FOV | 2.1 × 1.6 $D$ (211 × 160 mm) |
| Camera lens | AF-Nikkon Nikkor 35 mm |
| Focal length $f$ | 35 mm |
| Pixel size | 4.4 µm |
| Magnification factor $M$ | 0.035 |
| Image bit depth | 16 bits |
| Acquisiton frequency $f_s$ | 10 Hz (flow-on) |
| Nozzle inner diameter $D$ | 90 mm |
| Nozzle outer diameter | 100 mm |
| Lighting | 4 LED spotlights |

## 4 Experimental method

### 4.1 Experimental setup

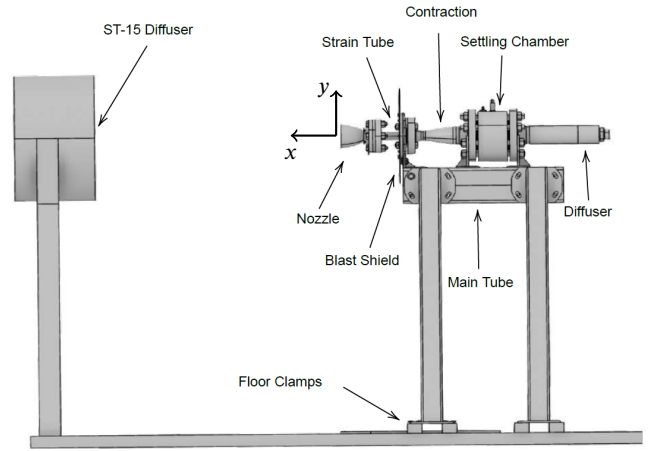

Figure 4: Schematic of the ASCENT test rig (adapted from de Kievit 2021)

Based on the considerations given in section 3 and guidelines given in Schwarz and Braukmann (2022) and taking inspiration from Nicolas et al. (2016) and Nicolas et al. (2017), an experimental setup was devised in the Aerodynamics High Speed Laboratory (HSL) of the Faculty of Aerospace Engineering of Delft University of Technology. The experimental setup consists of two main components: the ASCENT test rig and the circular camera-background array. The ASCENT test rig (de Kievit 2021), shown in fig. 4, houses the nozzle and is connected to a 300 m³ blow down tank which can provide a maximum pressure of 40 bar. The flow from the nozzle exhausts through a diffuser, also shown in the figure.

Around the ASCENT test rig, a circular camera-background array using X95 beams was designed, shown in fig. 5. It consists of four camera-background pairs, placed in a plane orthogonal to the nozzle x-axis, in a circular shape and attached to the beam structure. To provide sufficient illumination, four large LED spot lights were used. A summary of the optical parameters and equipment used is given in table 1.

The four cameras were placed in a 90° arc at equal 30° intervals, with each camera having its own speckle pattern background placed exactly opposite of it with respect to the nozzle axis (x-axis), as shown in fig. 6. The cameras and backgrounds had a target distance of $Z_A = 1000$ mm and $Z_D = 900$ mm to the nozzle axis, respectively. The choice of distances was mainly driven due to the desired field of view (FOV) in the plane of the nozzle of $2 \times 1.5\ D^2$. Table 2 shows the measured angular positions $\phi_i$ and distances $d_i$ to the nozzle x-axis of the cameras and backgrounds, as defined in figure fig. 6. A right hand coordinate system (shown in fig. 4, fig. 5 and fig. 6) is used. The x-axis points is aligned with the primary axis of the nozzle, the y-axis points up, and the z-axis is horizontal. The origin of the coordinate system is at the center of the nozzle exit. The angle $\phi$ is defined positive in the direction of the x-axis, and is measured from the negative z-axis. Note that the cameras and backgrounds are aligned with the y-z plane. Although not shown, the camera images use an r-axis (radius to nozzle center), which is always perpendicular to the camera axis and the x-axis.

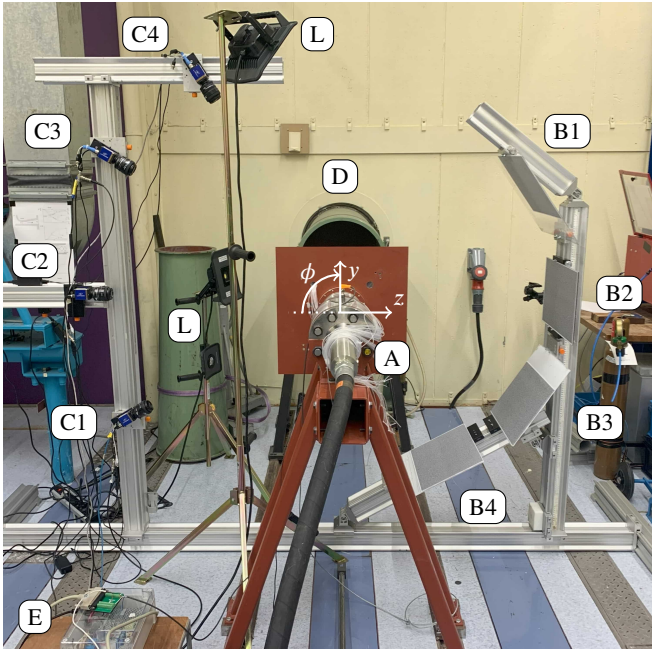

(a) Back view

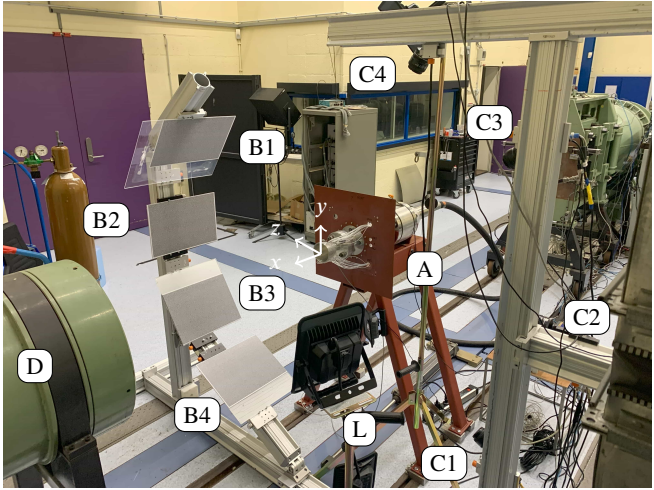

(b) Diagonal front view

Figure 5: Back and isometric front views of the experimental setup. The labels correspond to the following items: A - ASCENT test rig, B - backgrounds, C - cameras, D - diffuser, E - data acquisition control box, L - halogen lamps. The coordinate system is also given.

Table 2: Measured angular positions $\phi$ and distances $d$ to nozzle x-axis for each camera (C) and background (B) as defined in fig. 6. Here the distance between the cameras and nozzle axis is equivalent to $Z_A$. The distance between background and nozzle axis is equivalent to $Z_D$.

|    | $\phi$ [deg] | $d$ [mm] |    | $\phi$ [deg] | $d$ [mm] |
|----|---|---|----|---|---|
| C1 | -30.4 | 981 | B1 | 55.3 | 895 |
| C2 | 0.5 | 976 | B2 | 89.6 | 930 |
| C3 | 30.9 | 1001 | B3 | -57.8 | 930 |
| C4 | 57.7 | 946 | B4 | -29.3 | 885 |

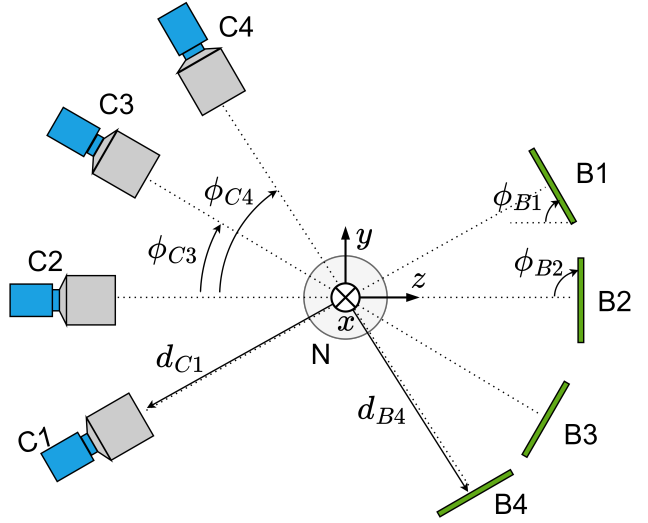

Figure 6: Schematic of experimental setup defining the angles and distances of table 2. Note that for conciseness not all angles and distances are shown. All angles drawn are positive in the direction shown. The letter correspond to the following items: B - backgrounds, C - cameras, N - nozzle.

The flow application of this study is a thrust-optimized parabola (TOP) contoured rocket nozzle (Ruf et al. 2009), of which a close up is shown in fig. 7. An overview of the geometric and flow parameters of the aluminium nozzle used in this study is given in table 3. This nozzle has an optimal expansion ratio at an NPR of 699.8, meaning the flow is highly overexpanded for all NPRs tested in this study. The NPR is the ratio of the pressure inside the settling chamber $p_c$ of the rocket nozzle to the ambient pressure $p_a$, NPR = $p_c/p_a$.

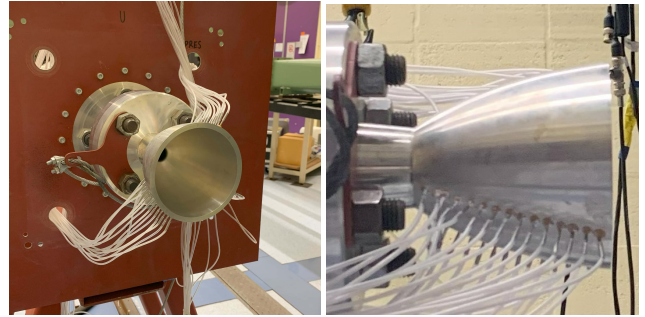

(a) Nozzle and backplate    (b) Nozzle side view

Figure 7: Aluminium nozzle used as flow application

Table 3: Nozzle properties

| Property | Value |
|---|---|
| Throat diameter | 16.35 mm |
| Exit diameter | 90 mm |
| Wall thickness | 5 mm |
| Length | 102.2 mm |
| Exit-to-throat area ratio | 30.29 |
| NPR at optimal expansion | 699.8 |

## 4.2 Data acquisition

To record the measurements, four computers were used, each connected to one of the cameras using Ethernet cables. During testing, each camera was operated using the Bobcat GEV-Player software, which allowed interfacing with the cameras and choosing various data acquisition parameters. Furthermore, during each test, one of the computers also collected NPR data at 20 Hz. A total of four commuters were used to facilitate the data stream of all four cameras.

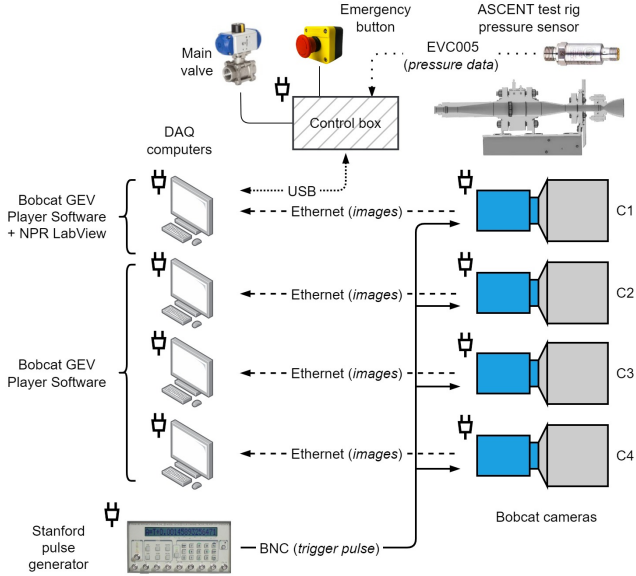

Figure 8: Schematic of the data acquisition setup. See de Kievit (2021) for more information about the control box.

In order to obtain instantaneous flow measurements from various angles, it is important for measurements of the different cameras to be taken at the same time. To achieve this camera synchronization, the cameras were connected using BNC cables to a digital pulse generator (Stanford Research Systems model DG535) which would send signals to simultaneously trigger all cameras at a desired measurement acquisition frequency.

To identify sets of 4 images of a snapshot which were incomplete due to computer errors, a simple algorithm was later devised that used the image time tags and the chosen acquisition frequency. The time stamps of each image were also used to isolate the ramp-up, steady, and ramp-down phases of each test as explained in section 4.5.

## 4.3 Background speckle pattern

Based on the chosen geometries of the experimental setup and camera parameters, the speckle-pattern backgrounds were designed. A small portion of a background panel used is shown in fig. 9 as an example. The speckle pattern was generated such that the size of the speckles corresponded to a size of 3-5 pixels per speckle as imaged by the cameras. Furthermore, the speckle density was iterated until an acceptable pattern with high enough spatial frequency but also enough contrast under the chosen optical conditions was obtained. The speckle patterns were printed on A3 paper and attached to rigid Plexiglas panels. Rigidity is important to avoid any potential vibrations of the background during testing, as it is of paramount importance for the background to remain as fixed in place as possible to obtain accurate measurements.

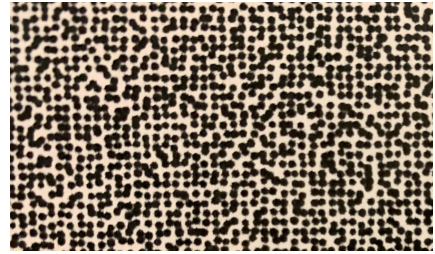

Figure 9: Sub-domain of the speckle background pattern used (not to scale)

## 4.4 Camera calibration

Before each test, the cameras were calibrated. This was necessary in order to obtain information about the positions of the cameras relative to each other, crucial for the tomographic reconstruction. Two methods were used for the calibration.

The first method consisted of recording a cylindrical grid placed at the nozzle, as shown in fig. 10a. The 80 mm diameter grid consisted of longitudinal and cylindrical lines with a spacing of 5 mm, which was aligned with the nozzle and visible from all cameras. These images then serve as a calibration images, and tomographic slices are later generated along each circular line of the cylinder. The cylinder is also used to vertically (in the FOV of the camera) align the measurements with respect to the x-axis of the nozzle (horizontal in the camera measurements).

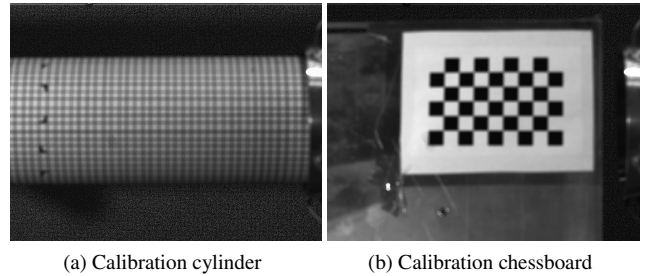

(a) Calibration cylinder      (b) Calibration chessboard

Figure 10: Calibration methods

The second method consisted of capturing images of a chessboard pattern (see fig. 10b) with known dimensions at different distances and orientations in the FOV of the four cameras. These images could then be used for multi-camera calibration using OpenCV (Deng et al. 2010) to extract the projection matrix of each camera, and in this way obtain more precise information on the location of the different cameras with respect to each other, focal length of the cameras and distance to the nozzle x-axis, to name a few parameters. Although not directly used in the present tomographic reconstruction, this information was recorded as it could in the future be used in a direct three-dimensional reconstruction from the two-dimensional projections, without first stacking the slices.

### 4.5 Testing procedure

To obtain data on the flow behaviour at the start-up and shut-down stages of nozzles jets, measurements of the flow were performed at various NPR. Three NPR values were chosen (12, 22 and 26) so that distinctly different flow regimes of the rocket nozzle could serve as test cases for inspecting the density fields. NPR values of 12 and 22 correspond to the so-called Free Shock Separation (FSS) state in which the flow separates from the nozzle wall, and forms a highly overexpanded supersonic plume (with trains of shock and expansion waves). For the highest NPR value of 26, the flow is in a Restricted Shock Separation (RSS) state, which is characterized by an annular supersonic plume (primarily attached to the nozzle wall) and a subsonic core region (Baars and Tinney 2013; Baars et al. 2015). Furthermore, several camera f-stops $f_{\#}$ and exposure times $t_{exp}$ were also tested. The test matrix summarizing the parameters for each test is given in table 4. It is important to note that all descriptions given in the following paragraphs are valid for each of the four cameras, and that all the data from each camera is kept separate until the reconstruction. Although not done in this study, different camera parameters were also measured to provide data and perform an assessment of their effects on the reconstruction. Test 1.4 (i.e. "22 hys 26") corresponds to a special test in which the NPR is first raised to 26 and then decreased and held steady at 22 to enable the flow to enter a hysteresis state). Test 3.4 (i.e. "slow 26") corresponds to a special test in which the NPR is increased to 26 much slower than over other tests, with the goal of capturing transient flow phenomena over this NPR increase.

Table 4: Test matrix showcasing different combinations of parameters used for the different tests.

| Test | NPR [-] | $t_{exp}$ [ms] | $f_{\#}$ | Runs |
|---|---|---|---|---|
| 1.1 | 12 | 50 | f/5.6 | 3 |
| 1.2 | 22 | 50 | f/5.6 | 3 |
| 1.3 | 26 | 50 | f/5.6 | 3 |
| 1.4 | 22 hys 26 | 50 | f/5.6 | 2 |
| 2.1 | 12 | 250 | f/5.6 | 1 |
| 2.2 | 22 | 250 | f/5.6 | 1 |
| 2.3 | 26 | 250 | f/5.6 | 1 |
| 3.1 | 12 | 50 | f/2.8 | 1 |
| 3.2 | 22 | 50 | f/2.8 | 1 |
| 3.3 | 26 | 50 | f/2.8 | 1 |
| 3.4 | slow 26 | 50 | f/2.8 | 1 |

Each test started with the recording of approximately 100 "flow off" images. The average of these images was later on used as the "flow off" image for the cross-correlation. The reason for this is that a single image contained too much noise due to the low exposure time of the cameras.

Once, the flow off images were recorded and the safety procedures preceding the test performed, the "flow on" images were recorded. To do this, the pulse generator was turned on, and the simultaneous recording of the four cameras and logging of the NPR was started. The first phase consisted of slowly opening the main valve to achieve a linear NPR ramp-up until the target NPR was attained. The flow was then kept steady at this target NPR for approximately 20 seconds or until 200 images were recorded, after which the valve was slowly closed and the NPR ramped down. The NPR measurements for the tests 3.1, 3.2 and 3.3 is shown in fig. 11. Note that all three phases were measured in order to fully characterize the nozzle jet flow structure from start-up to shut-down. In total, approximately 340 GB of images were recorded.

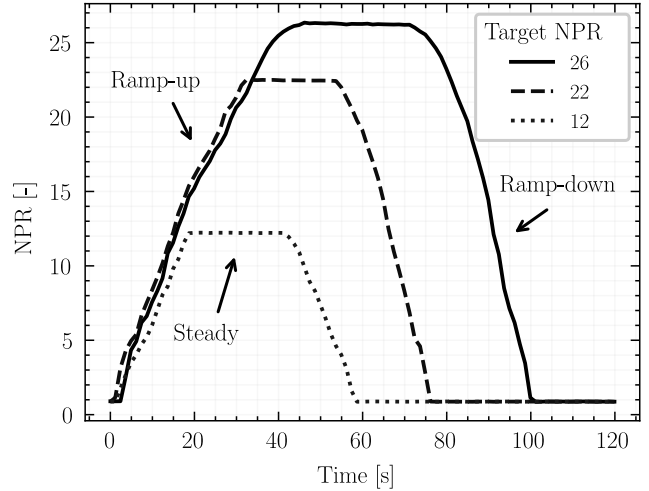

Figure 11: Time evolution of NPR for tests 3.1, 3.2 and 3.3

## 5 Data processing

### 5.1 BOS data processing

From all the data collected, the data subset corresponding to tests 3.1, 3.2 and 3.3 (i.e. $t_{exp} = 50$ ms and $f_{\#} = 2.8$, see table 4) was selected for further processing. Furthermore, only the steady phase of the tests was used. The results were obtained only for this data set, as not all data could be processed due to the volume and because pre-processing of this subset showed good resolution and relatively high SNR. An example of a flow-off and flow-on image is shown in fig. 12a and fig. 12b, respectively. The images are zoomed-in around half of the nozzle in fig. 12c and fig. 12d to show the strong density gradients at the boundary of the jet. Although the images had low light intensity, the intensity value difference between the black and white pixels of the background pattern still surpassed 100 counts, enough for proper distinguishing between black dots and white background. Also, the darkest pixels were still above 50 counts, meaning no saturation of the dark pixels' intensity was present. Defect pixels were dealt with by averaging the surrounding pixel values.

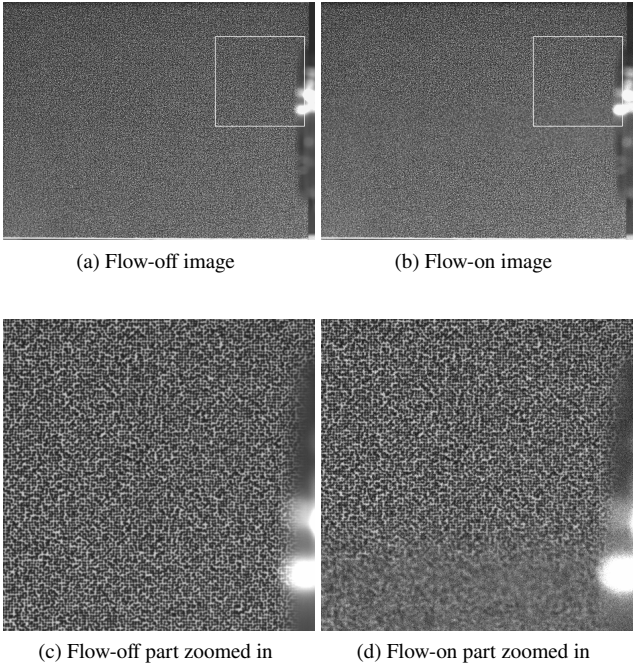

(a) Flow-off image  (b) Flow-on image

(c) Flow-off part zoomed in  (d) Flow-on part zoomed in

Figure 12: Flow-off and flow-on images at NPR $\approx$ 22. The contrast of the images has been increased, and the white boxes correspond to the part that is zoomed-in.

The flow-off images used in the cross-correlation are the average of approximately 100 flow-off images. This averaging increases the contrast and quality of the flow-off image. To obtain the image displacements, approximately 50-100 flow-on images of the steady phase (i.e. at a constant NPR) were cross-correlated with the flow-off image to obtain the image displacement $\Delta \vec{x}$ (at each point in the image) based on the IWs chosen.

To increase the accuracy of the cross-correlation, an iterative approach was used. This is necessary as strong density gradients due to strong shear flows are present, leading to optical blurring. This has a large negative effect on the cross-correlation as the background pattern is not only blurred, but also deformed. To alleviate these effects, the approach uses multiple passes (in this case 4) of the cross-correlation with decreasing window sizes, as well as adaptive window deformation to increase the robustness of the displacements estimation (Scarano 2002). Additionally, in between passes a validation scheme is implemented which detects and replaces outliers based on a median test with a two standard deviation cut-off (Westerweel and Scarano 2005). Finally, to achieve sub-pixel resolution, the peak is interpolated using a sinc function. The details on the cross-correlation parameters used is given in table 5.

To evaluate the quality of the displacements, the SNR of the correlations is also computed. Here, the SNR is defined as the ratio between the two highest peaks in the correlation map, i.e. SNR = $\gamma_1/\gamma_2$. A SNR higher than 1.5 was deemed of sufficient quality.

Table 5: Cross-correlation settings

| Setting | Description |
| --- | --- |
| Number of images | $\sim$100 ($\times$4 cameras) |
| Image resolution | $1624 \times 1236$ pix$^2$ |
| Number of passes | 4 |
| FOV | $2.1 \times 1.6\ D^2$ |
| Vector spacing | 0.67 mm |
| Initial window size | $63 \times 63$ pix$^2$ |
| Final window size | $21 \times 21$ pix$^2$ ($2.72 \times 2.72$ mm$^2$) |
| Final image size | $316 \times 239$ pix$^2$ |
| Image resolution | 0.13 mm/pix |
| Window overlap | 75% |

Once all flow-on images had been processed to obtain the displacements at each location in the image, the displacement fields were averaged to obtain an average displacement field, corresponding to a time-averaged scenario instead of instantaneous snapshots. The displacement magnitude $\|\Delta \vec{x}\| = \sqrt{\Delta x^2 + \Delta y^2}$ is then directly proportional to the line-of-sight integrated density gradients.

## 5.2 Poisson solution and tomographic reconstruction

To obtain the line-of-sight integrated density field, the Poisson equation given by eq. (14) has to be solved. The source term, given by eq. (15), is computed by taking the dot product of the gradient operator and the displacement vector, i.e. $\nabla \cdot \Delta \vec{x} = \frac{\partial \Delta x}{\partial x} + \frac{\partial \Delta y}{\partial y}$. This is numerically implemented by using a finite difference scheme to approximate the derivatives of $\Delta x$ and $\Delta y$. This is then scaled by the factor given in eq. (15) to obtain the source term of the Poisson equation. Solving the Poisson equation leads to the line-of-sight integrated density. This is done using a simple (second order) central difference scheme in both $x$ and $y$ directions. Neumann boundary conditions are used at the left and right boundary to specify that the gradient of the density orthogonal to and at these boundaries is zero, i.e. $\partial \rho^*/\partial x = \partial \rho^*/\partial y = 0$. For the top and bottom boundaries, Dirichlet conditions are used. These specify that the values of the argument of the Poisson equation is known, and here was set to product of the ambient density and the distance between camera and background, i.e. $\rho^* = \rho_0 Z_B$.

The line-of-sight integrated density field can then be back-projected to obtain the three-dimensional density field. In this study, vertical lines orthogonal to the x-axis are taken and back-projected to obtain two-dimensional slices of the density field. These slices are then stacked, obtaining a (quasi-) three-dimensional reconstruction of the density field. The numerical implementation of the back-projection is performed by generating a sinogram of the projections (a field of the projections for each angle) and using an inverse radon transform available in the Scikit python library. To improve the reconstruction, initially a filtered back-projection was used as a first guess in the simultaneous algebraic reconstruction technique (SART) (see Andersen and Kak 1984 for more details). However, better results with less tomographic artefacts were obtained us-ing only the SART algorithm, and thus one iteration of this technique was used to reconstruct slices. The reconstruction was performed for each vertical line in the images. The obtained density field was then normalized with respect to the ambient density. In this work, the calibration step itself was omitted. This did not negatively affect the qualitative nature of the results (see section 6.3), and could be implemented in future work to improve the reconstruction.

# 6 Results and discussion

## 6.1 BOS displacements

After processing the data, results for tests 3.1, 3.2 and 3.3 (i.e. $t_{exp} = 50$ ms and $f_\# = f/2.8$, see table 4) are shown in this section. The NPR measurements for these tests are shown in fig. 11, from which it can be seen that the NPR remains relatively constant at the target NPR during the steady phase. The time-averaged image displacements for these tests are shown in fig. 14. The columns correspond to the image displacements $\Delta x$, $\Delta y$ and the magnitude of the image displacements $\|\Delta \vec{x}\|$ (from left to right, respectively). The rows correspond to an NPR of 12.2, 22.5 and 26.2 (from top to bottom, respectively). The nozzle is not shown, and these figures show the flow that comes out of the nozzle, flowing from left to right. The axis have been normalized with respect to the nozzle diameter. From these results, the shock structures are clearly visible and match the ones obtained by de Kievit (2021). Also, the time-averaged flow structure appears to be quite symmetrical.

The unsteady nature of the flow can be deduced by looking at fig. 15, which shows the standard deviation at each location in the time-averaged images. A snapshot of the instantaneous displacement field at a NPR of 22.5 is shown in fig. 13, from which the unsteady and turbulent nature of the flow can also be seen. In this figure, the image displacements are shown as a vector field, i.e. $\Delta \vec{x}$ at each point in the image. Note that the results are shown only for camera 3 since the general aspects of the results are similar for all cameras.

## 6.2 Poisson equation solution

As described in section 3 and section 5, the source term of the Poisson equation is computed. The source terms for the Poisson equations of camera 3 (which uses the displacements given in fig. 14), are given in fig. 16. The increased granularity in these images shows that the gradient operator adds noise to the images.

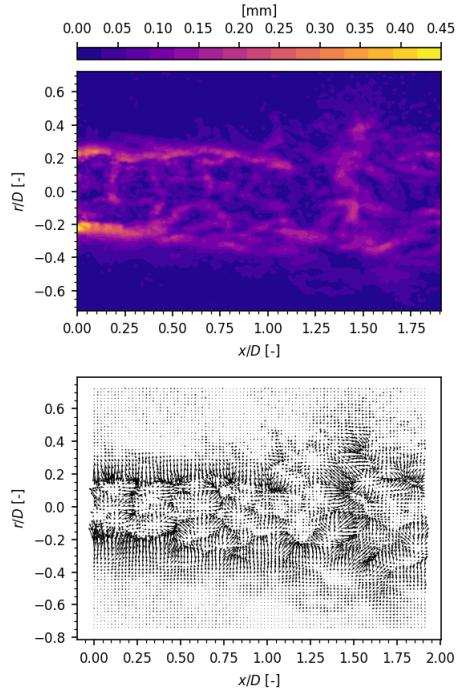

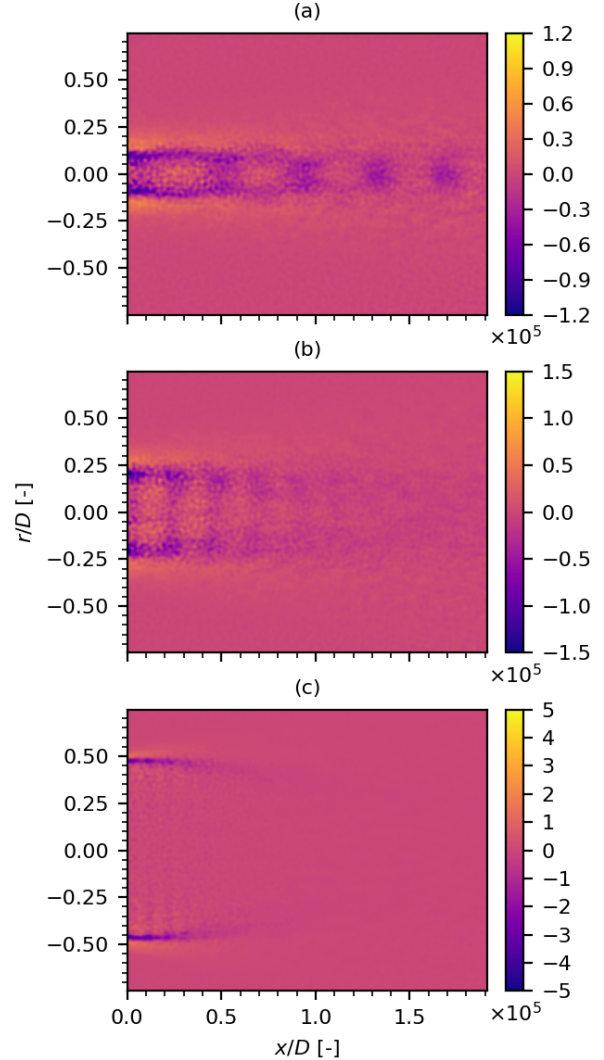

Figure 13: (a) Displacement vector $\Delta \vec{x}$ field. For better visibility, the vector lengths have been scaled by a factor 6.5 and (b) Displacement magnitude contour plot.

Figure 16: Source term used in the Poisson equation [kg/m$^4$] for NPR 12.2 (a), 22.5 (b) and 26.2 (c) of camera 3.

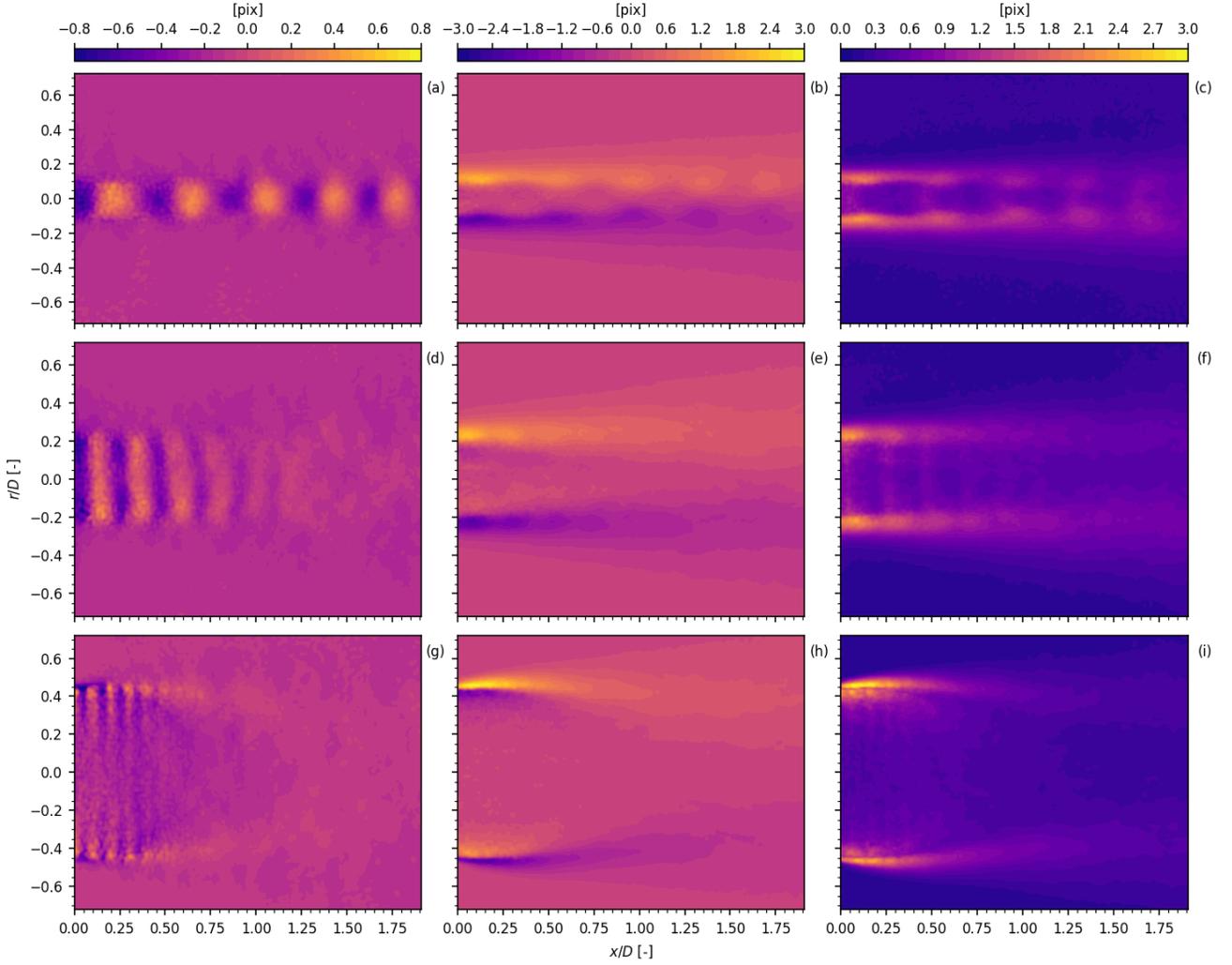

Figure 14: Time-averaged image displacements for tests 3.1, 3.2 and 3.3 ($t_{exp}$ = 50 ms and $f_\#$ = $f$/2.8). The top row (subfigures a-c), middle row (subfigures d-f) and bottom row (subfigures g-i) correspond to a NPR of 12.2, 22.5 and 26.2, respectively. The left (subfigures a, d, g), middle (subfigures b, e, h) and right (subfigures c, f, i) column corrspond to the x-direction displacement $\Delta x$, y-direction displacement $\Delta y$, and image displacement magnitude $||\Delta \vec{x}||$, respectively.

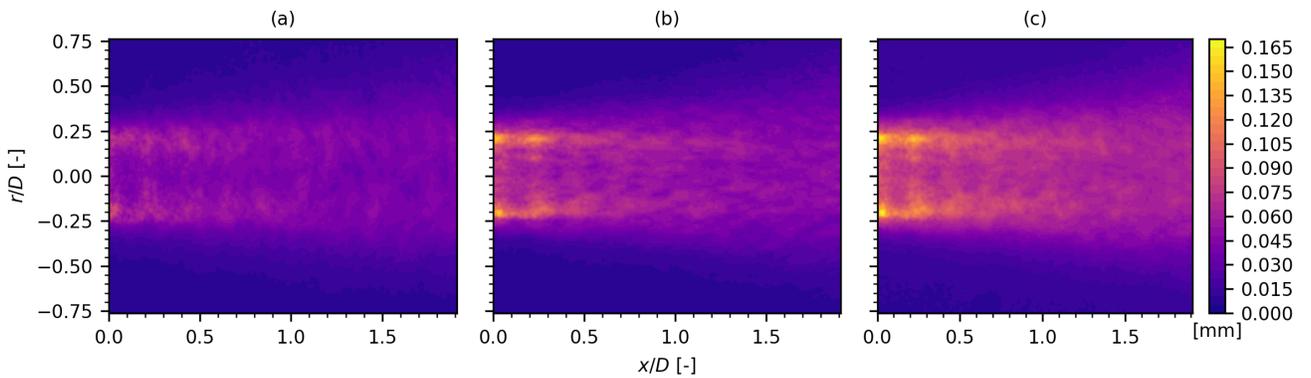

Figure 15: From left to right: standard deviation of $\Delta x$, standard deviation of $\Delta y$, and norm of the $\Delta x$ and $\Delta y$ standard deviations.

## 6.3 Tomographic reconstruction

With the line-of-sight integrated density fields obtained, the tomographic reconstruction can be performed. Here, two tomographic reconstructions are demonstrated. The first one assumes that we have many more projections around the flow, in this case 18 cameras equally spaced in azimuth to cover the full 360° viewing angle. This is to demonstrate that, in the case where more viewing angles would be present, the reconstruction can vastly be improved. The reasoning for doing this is that by taking a time-averaged flow, the projections at different angles should be similar (since the nozzle is symmetric). Thus, if a reconstruction using the same data for all these "virtual" projections is attempted, a good approximation to the time-averaged density field could be obtained (assuming the camera axis is perfectly perpendicular to the flow). The second reconstruction corresponds to the one of the experimental setup, i.e. using four cameras at 30° inter-camera spacing.

To obtain the three-dimensional density field, we first back-project a single vertical (in the image plane) line of the line-of-sight integrated density distribution from many angles to obtain a two-dimensional slice of density orthogonal to the nozzle axis (x-axis). As an example, the line at $x/D = 0.05$ is chosen and back-projected to obtain the two-dimensional slice of density shown in fig. 18 and fig. 19 for the eighteen projections case and the four projections case, respectively. The density in these slices is normalized with respect to the ambient flow density $\rho_0$. From these figures reconstruction artefacts can be seen, especially at the boundaries. This is an inherent problem of tomographic reconstruction using few projections. Also, note how the density field in the four projections case appears to be slightly oval shaped. This could be due to only having a 90° coverage in azimuth instead of 360°, as for the eighteen projections case. Furthermore, the NPR 26 case is poorly reconstructed in the case of only four projections.

This procedure is repeated for all lines in $x/D$ and the slices are stacked, yielding a (quasi-) three-dimensional density field, the $x/D - z/D$ plane of which is shown in fig. 20 and fig. 21 (for the eighteen and four projections cases, respectively). This plane was selected as it was not a plane of one of the cameras and is obtained purely from the reconstruction, making it a good indicator of the reconstruction quality. From these results, the density field appears to be quite symmetric about the $x/D$ axis. Also, for both the eighteen and four projections cases, the shock diamonds are clearly visible. However, the density fields lack sharpness in the reconstructed density fields, as normal and oblique shockwaves typically present in such a flow are not clearly visible. It is unclear if this is due to the reconstruction technique or the experimental parameters chosen. Furthermore, the flow is much sharper in $z/D$ direction for the eighteen projections case, where a clear boundary with respect to the ambient air can be seen. fig. 22 shows the normalized density on the center line ($z/D = 0$) for the two-dimensional density slices shown in fig. 20 and fig. 21. From these, the shock spacing can be extracted: the shock spacing for NPR = 12.2 is approximately 0.4 $D$, and the shock spacing for NPR = 22.5 is approximately 0.25 $D$.

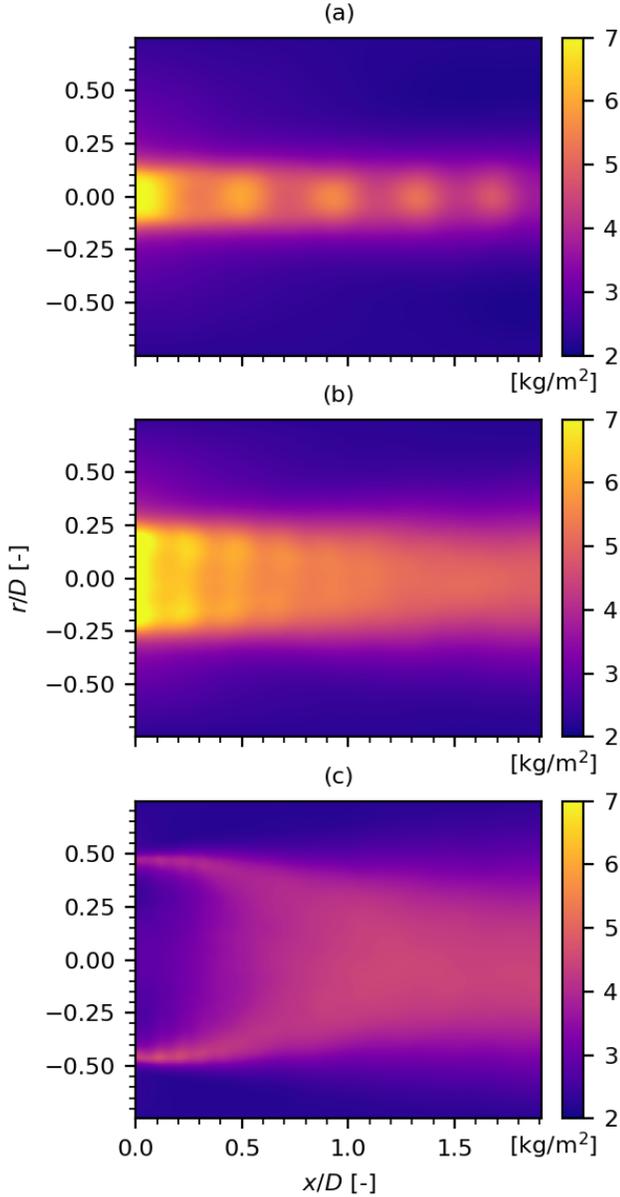

Figure 17: Line-of-sight integrated density field $\rho^*(x, y)$ for NPR 12.2 (a), 22.5 (b) and 26.2 (c) of camera 3.

The solution to the Poisson problem using these source terms leads to the line-of-sight integrated density field $\rho^*(x, y)$, given in fig. 17. As can be seen, solving the Poisson equation tends to apply a smearing out/ blurring effect, and further increases the inability to resolve sharp density gradients. The whole procedure of obtaining the Poisson solution can be repeated for the different cameras which yields similar results, mainly because the process is applied to the time-averaged images instead of the instantaneous ones. For conciseness, the results of the other cameras are not shown.

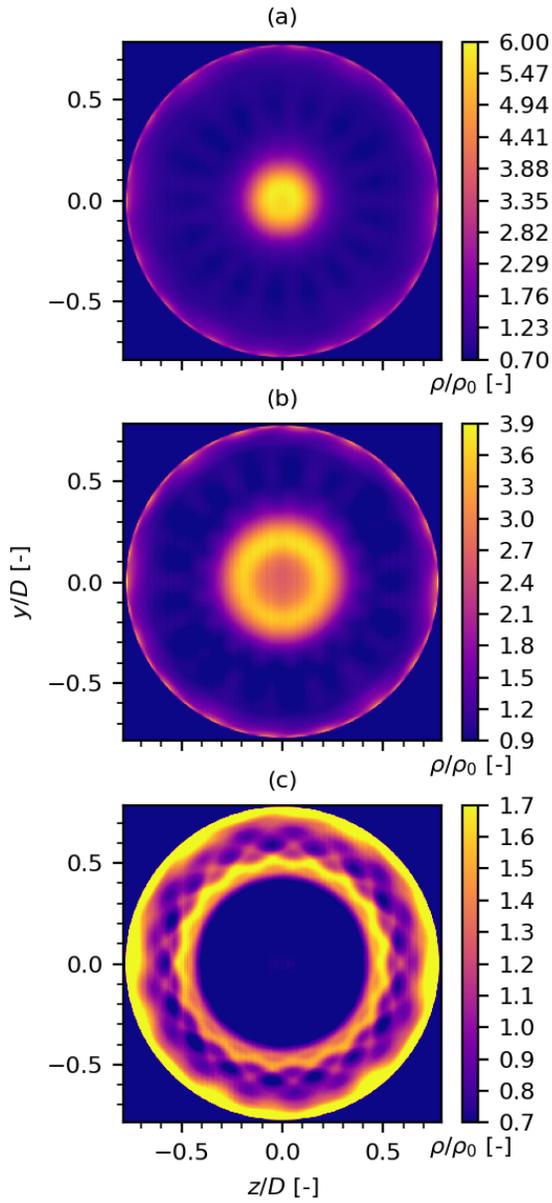

Figure 18: Normalized density field $\rho/\rho_0$ slice for NPR 12.2 (a), 22.5 (b) and 26.2 (c) of camera 3 at $x/D = 0.05$ for the eighteen projections case. Note the tomographic reconstruction artefacts at the circular boundaries, especially for (c).

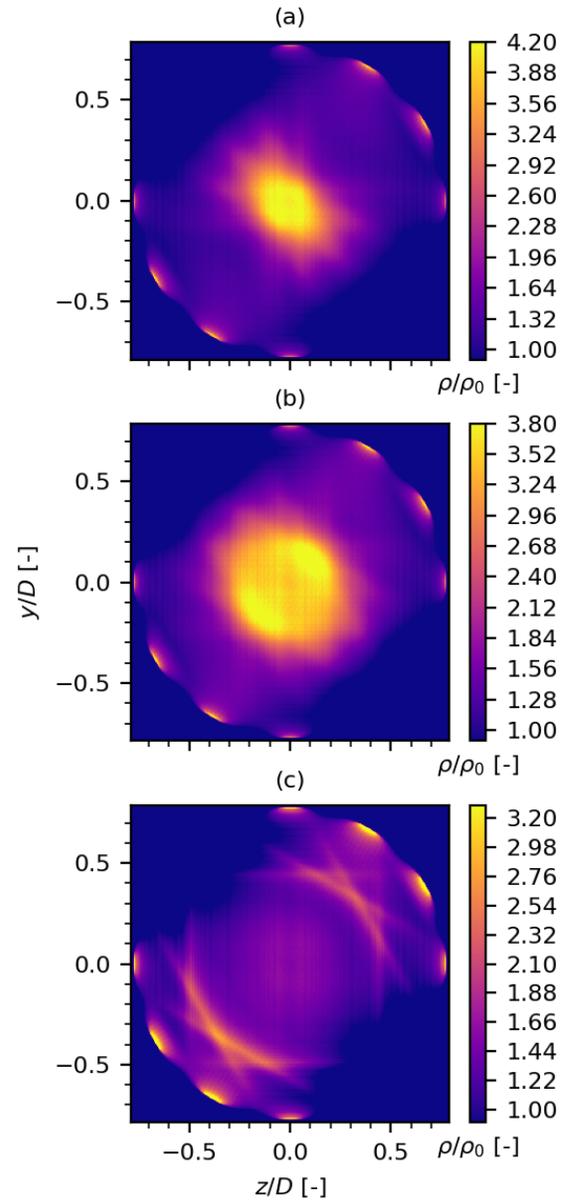

Figure 19: Normalized density field $\rho/\rho_0$ slice for NPR 12.2 (a), 22.5 (b) and 26.2 (c) of camera 3 at $x/D = 0.05$ for the four projections case. Note the tomographic reconstruction artefacts at the circular boundaries, especially for (c). This reconstruction is noisier than the eighteen projections case shown in fig. 18

Finally, normalized density iso-surfaces of the full three-dimensional density fields are shown in fig. 23 and fig. 24 for a NPR of 12 and 22 of the eighteen projections case, respectively. From these, the three-dimensional structure of the flow can be observed. Density variations typical of shock diamonds are visible. The density field counter part for the four projections case is shown in fig. 26 and fig. 27. Here, although the three-dimensional density iso-surfaces are less smooth, a three-dimensional structure with diamonds is still visible albeit with more tomographic artefacts and noise.

The NPR 26 case for the eighteen and four projections case is shown in fig. 25 and fig. 28, respectively. For the eighteen projections case, the general flow characteristics are somewhat visible although noise is present. This NPR is reconstructed much less clearly for the four camera case, where it is almost not possible to extract an apparent flow structure. This could, however, be due to the way of visualizing the three-dimensional iso-surfaces, since the slice shown in fig. 21 (c) appears to be smoother.

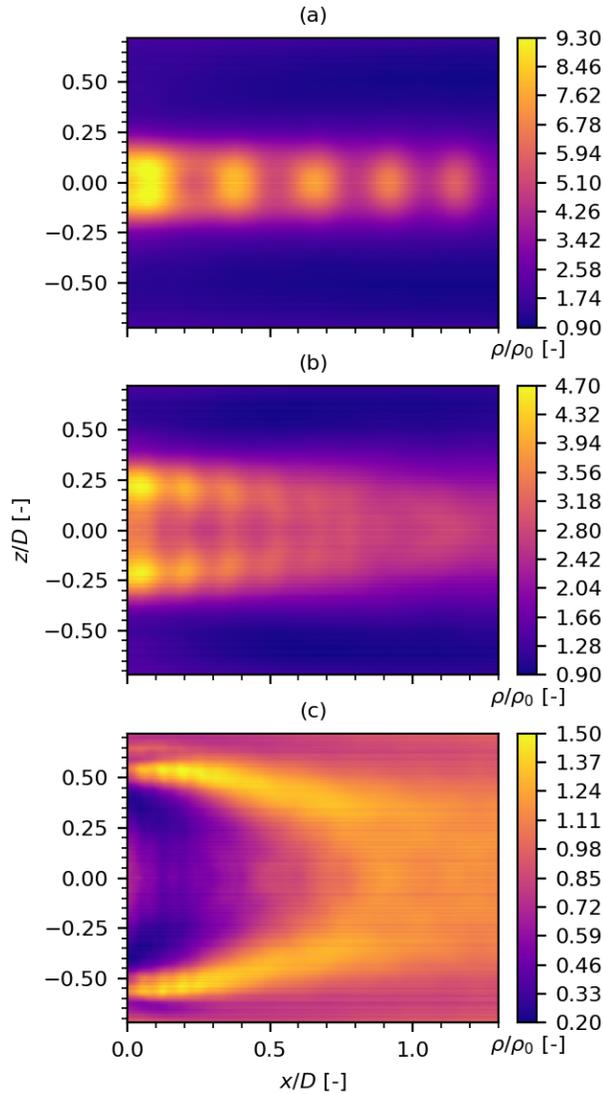

Figure 20: $y/D = 0$ plane of the reconstructed (normalized) density field for NPR 12.2 (subfigure a), 22.5 (subfigure b) and 26.2 (subfigure c) for the eighteen projections case.

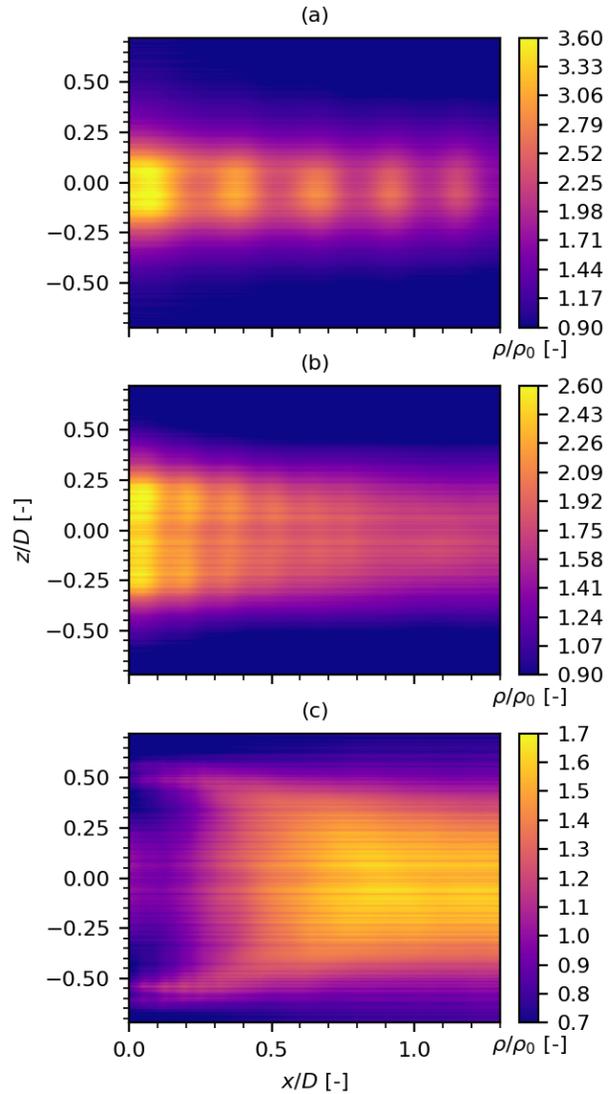

Figure 21: $y/D = 0$ plane of the reconstructed (normalized) density field for NPR 12.2 (subfigure a), 22.5 (subfigure b) and 26.2 (subfigure c) for the four projections case.

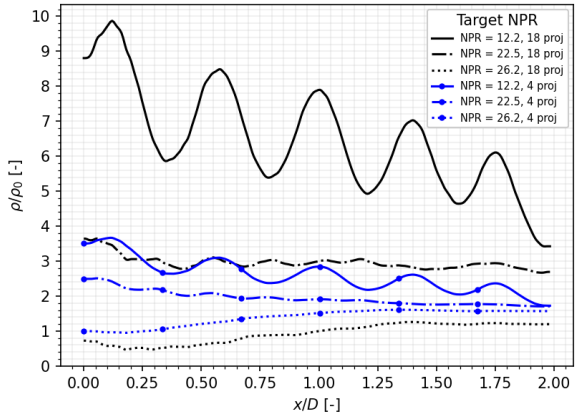

Figure 22: Normalized density profiles at $y/D = 0$ and $z/D = 0$ for the various reconstructed three-dimensional density fields.

Time restrictions in the project led to some of the results not being of the desired quality, allowing ample opportunities for improvement. First of all, a validation could not be performed. Then, four cameras could not be used and instead all the data used comes from camera 3. This is not expected to have a large effects on the time-averaged results as explained earlier. However, it might have a large effect on the instantaneous tomographic reconstruction (not performed in this study). Furthermore, there are some significant differences in the values obtained for the eighteen and four projections case, the source of which is currently unknown. Furthermore, although the reconstructions yield acceptable results using few projections, the calibration was not used, and instead the data from camera 3 was manually centered. Finally, although it is possible to perform the tomographic reconstruction of instantaneous snapshots using the data from multiple cameras, the results were only generated for the time-averaged flows.

## 7 Concluding remarks

This study presented the development of a tomographic background-oriented Schlieren setup and data processing technique applied to an overexpanded nozzle jet. An experimental setup was devised using four cameras, and the BOS results were used as a source term to a Poisson equation, the solving of which yielded two-dimensional line-of-sight integrated density fields. Individual vertical lines in the camera plane of this field were then back-projected to reconstruct the two-dimensional slices of the density field orthogonal to the nozzle axis. The slices were then stacked, resulting in a three-dimensional density field.

The results show that the method implemented is able to obtain the BOS displacements accurately. The main flow characteristics such as shock diamonds are visible, even with the inherent limited spatial resolution of BOS. Regarding the tomographic reconstruction, although most three-dimensional flow characteristics are visible, some improvements are required to reach the level of reconstructions found in literature. First, it is recommended to improve the experimental setup by implementing a proper calibration of the cameras. It is also recommended to implement the more advanced methods by Grauer et al. (2018), Nicolas et al. (2017) and Amjad et al. (2020) which are computationally more demanding but do not make the parallel rays assumption, and appear to produce better results. Furthermore, use should be made of the data provided at different camera parameters to investigate the parameter combination leading to the sharpest reconstructions. Finally, it is recommended to perform a validation of the obtained results in this study. This could be done by comparing the present results with the Schlieren and PIV Mach contour results obtained by de Kievit (2021) for the same nozzle.

## Supplementary material

For access to the data, please contact the authors. Access to the code and data processing scripts can be found at the following link: https://github.com/joabron/TBOS-aerodynamics.

## Acknowledgements

The author would like to thank Martina Formidabile, Alessandro D'Aguanno and Adrian Grille Guerra for their help during the experimental campaign and for the numerous helpful discussions. The author also thanks Adam Head and Peter Duyndam for providing technical support during the experimental campaign.

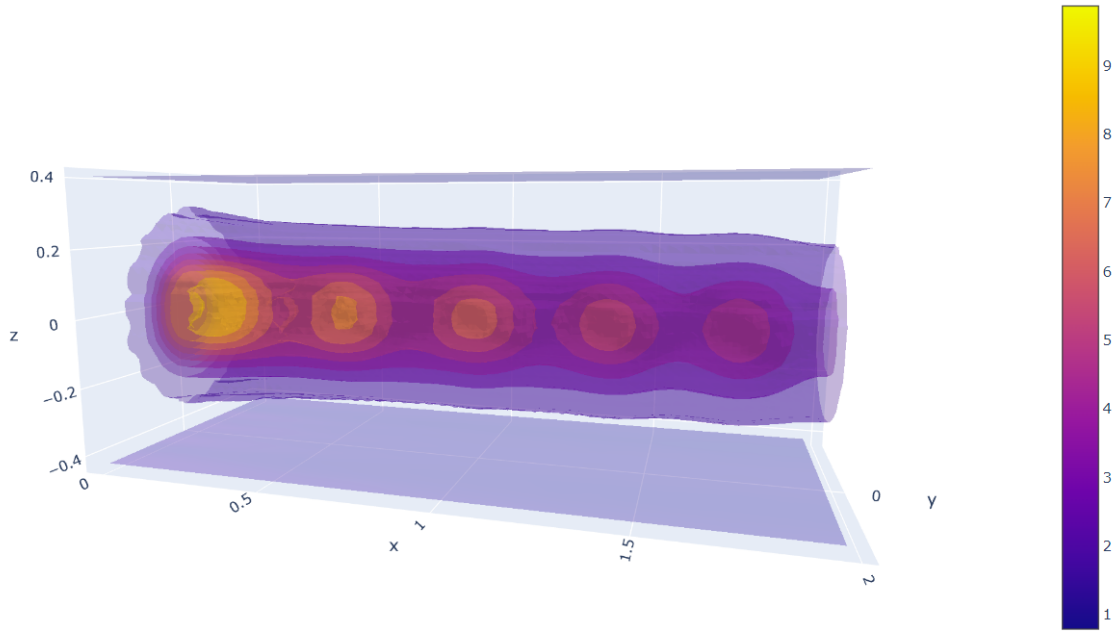

Figure 23: Three-dimensional reconstructed normalized density field $\rho/\rho_0$ [-] for NPR 12.2 for the eighteen projections case (using only camera 3). The axes are non-dimensionalized with respect to the nozzle diameter $D$.

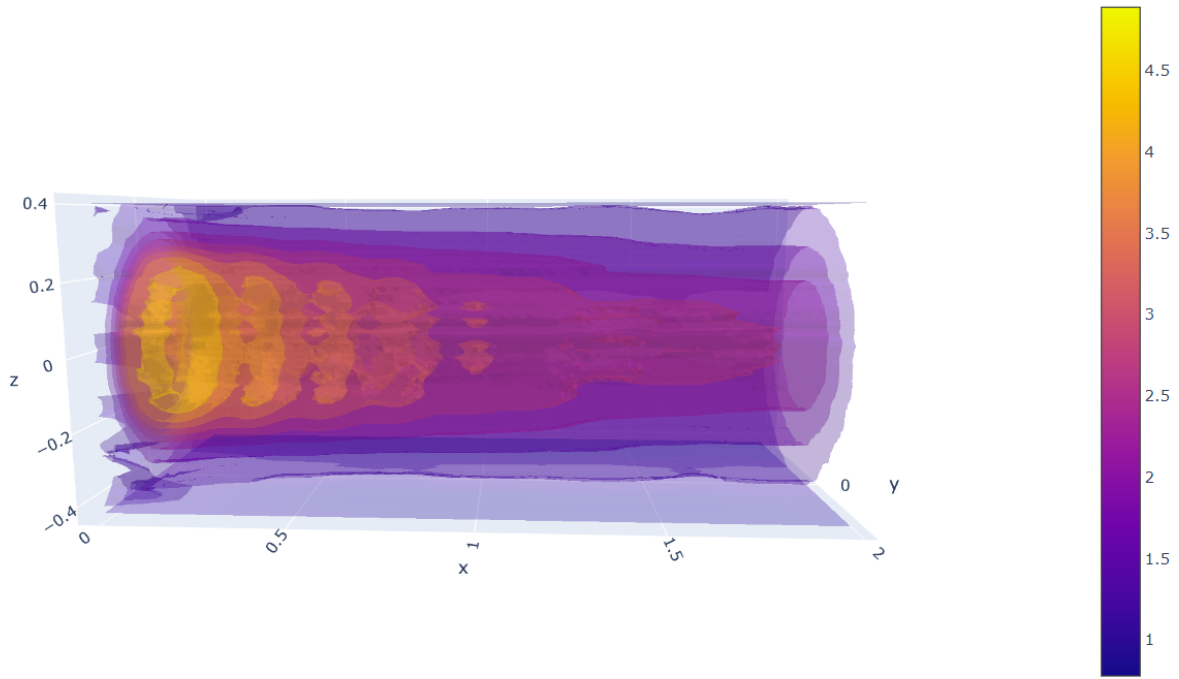

Figure 24: Three-dimensional reconstructed normalized density field $\rho/\rho_0$ [-] for NPR 22.5 for the eighteen projections case (using only camera 3). The axes are non-dimensionalized with respect to the nozzle diameter $D$.

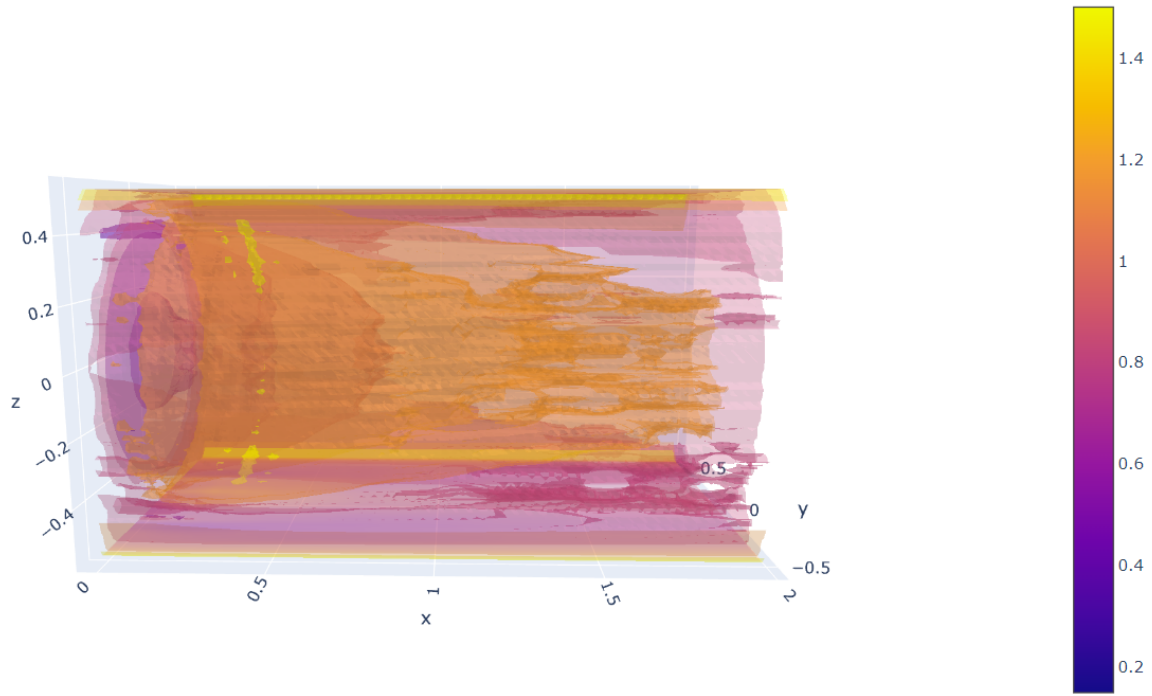

Figure 25: Three-dimensional reconstructed normalized density field $\rho/\rho_0$ [-] for NPR 26.2 for the eighteen projections case (using only camera 3). The axes are non-dimensionalized with respect to the nozzle diameter $D$.

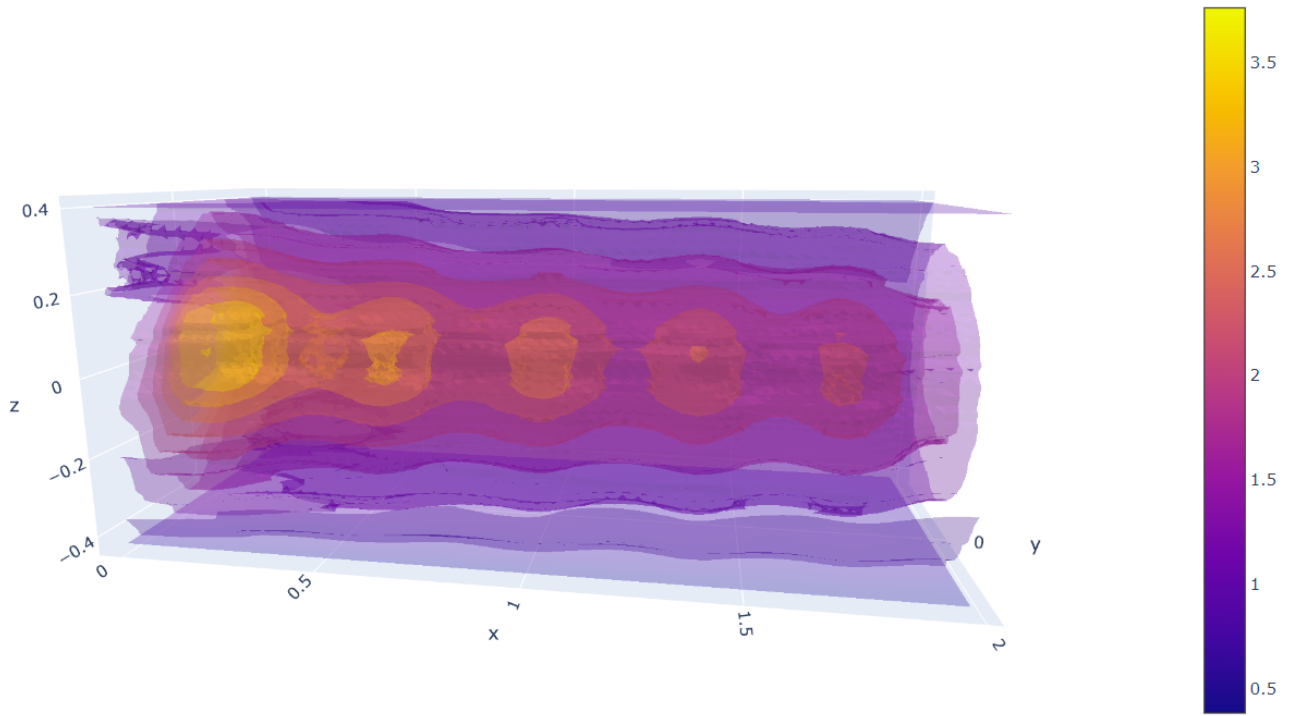

Figure 26: Three-dimensional reconstructed normalized density field $\rho/\rho_0$ [-] for NPR 12.2 for the four projections case (using only camera 3). The axes are non-dimensionalized with respect to the nozzle diameter $D$.

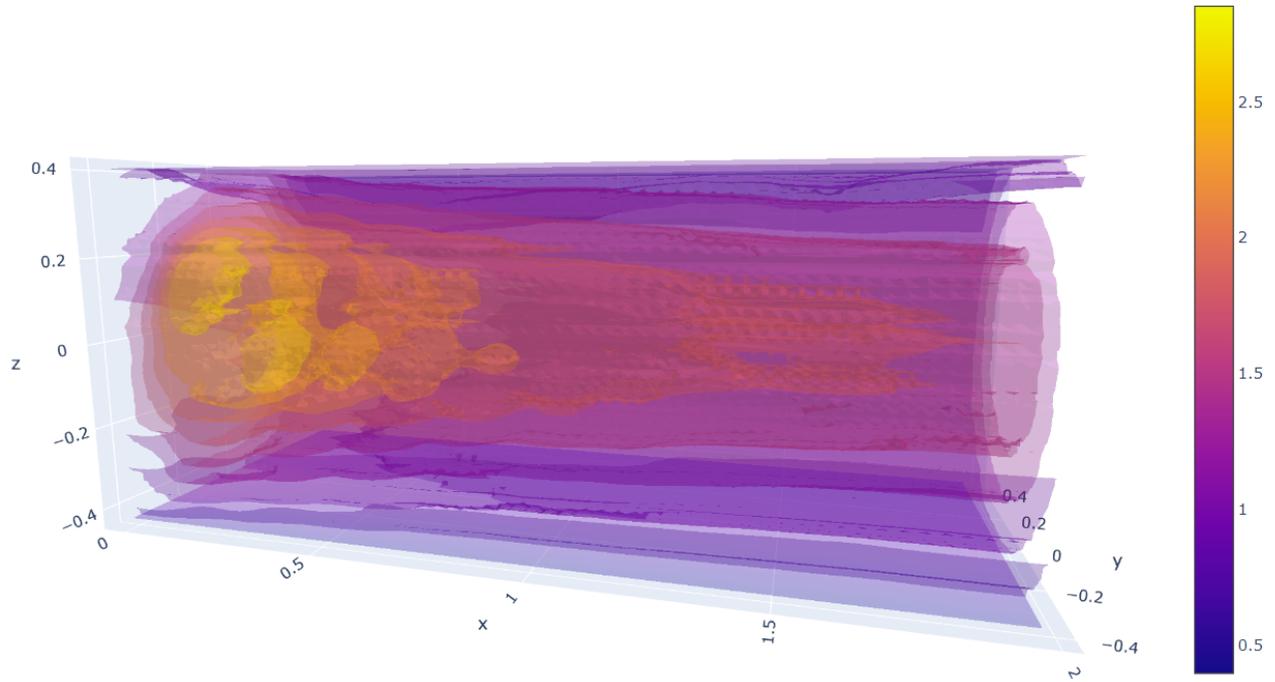

Figure 27: Three-dimensional reconstructed normalized density field $\rho/\rho_0$ [-] for NPR 22.5 for the four projections case (using only camera 3). The axes are non-dimensionalized with respect to the nozzle diameter $D$.

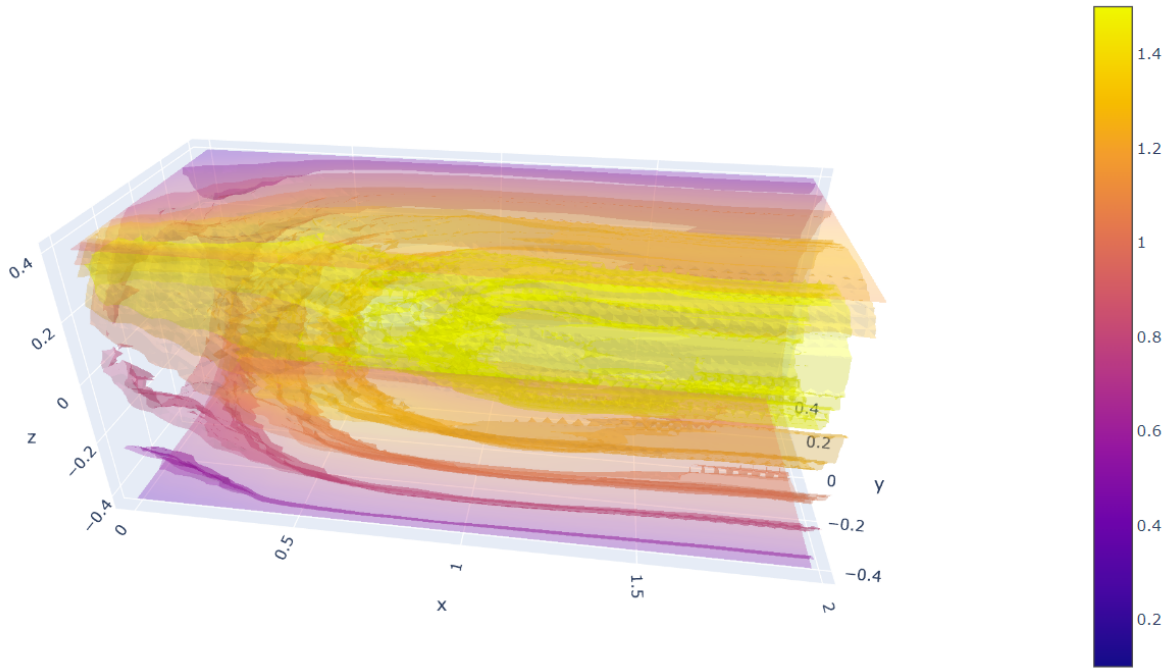

Figure 28: Three-dimensional reconstructed normalized density field $\rho/\rho_0$ [-] for NPR 26.2 for the four projections case (using only camera 3). The axes are non-dimensionalized with respect to the nozzle diameter $D$.